\documentclass[%
 reprint,
 amsmath,amssymb,
 aps,
]{revtex4-1}

\usepackage{graphicx}
\usepackage{dcolumn}
\usepackage{bm}

\begin{document}

\preprint{APS/123-QED}

\title{Dynamics of Panic Pedestrians in Evacuation}
\thanks{Dynamics of Panic pedestrians}%

\author{Dongmei Shi}
\email{sdm21@mail.ustc.edu.cn}

 \affiliation{Department of Physics, Bohai University, Jinzhou Liaoning, 121000, P. R. C}
 \author{Wenyao Zhang}%
\affiliation{%
 Department of Physics, University of Fribourg, Chemin du Musee 3, CH-1700 Fribourg, Switzerland.
 }%
 \author{Binghong Wang}%
 \affiliation{%
 Department of Modern Physics, University of Science and Technology of China, 230026, Hefei Anhui, P.R.C
 }%

\date{\today}

\begin{abstract}
A modified lattice gas model is proposed to study pedestrian evacuation from a single room. The payoff matrix in this model represents the complicated interactions between selfish individuals, and the mean force imposed on an individual is given by considering the impacts of neighborhood payoff, walls, and defector herding. Each passer-by moves to his selected location according to the Fermi function, and the average velocity of pedestrian flow is defined as a function of the motion rule. Two pedestrian types are included: cooperators, who adhere to the evacuation instructions; and defectors, who ignore the rules and act individually. It is observed that the escape time increases with the panic level, and the system remains smooth for a low panic level, but exhibits three stages for a high panic level. We prove that the panic level determines the dynamics of this system, and the initial density of cooperators has a negligible impact. The system experiences three phases, a single phase of cooperator, a mixed two-phase pedestrian, and a single phase of defector sequentially as the panic level upgrades. The phase transition has been proven basically robust to the changes of empty site contribution, wall's pressure, and noise amplitude in the motion rule. It is further shown that pedestrians derive the greatest benefit from overall cooperation, but are trapped in the worst situation if they are all defectors.
\begin{description}

\item[PACS numbers]
87. 23. Ge,  02. 50. Le, 87. 23. Cc, 89. 90. +n

\end{description}
\end{abstract}

\maketitle


\section{\label{sec:level1}introduction}

Evacuation of pedestrians under panic has been extensively studied \cite{model1,model2,model3,model4,model5,model6,model7,model8,model9,model10, model11, model12, model13, model14, model15, model16, model17, model18, model19, ca}, because an understanding of the dynamic features of this phenomenon can reduce the incidence of injury and death. Escape panic can occur in different confinement sizes ranging from a rioting crowd in a packed stadium to stunned customers in a bar. Escape panic is characterized by strong contact interactions between selfish individuals that quickly give rise to herding, stampede, and clogging \cite{feature1, feature2, feature3, feature4, feature5,feature6, feature7, feature8, feature9, feature10, feature11}. Experimental observations and numerical simulations are two approaches that have been widely applied to study this issue. Simulation results have revealed some interesting dynamic features, such as pedestrian arch formation around the exit, herding, and interference between arches in multiple-exit rooms \cite{model2}. Disruptive interference, self-organized queuing, and scale-free escape dynamics \cite{feature6, feature7} have also been observed. Experiments in genuine escape panic are difficult to set up because of ethical and legal concerns.

However, escape panic can be simulated by solving a set of coupled differential equations \cite{model2,model3} or by applying the cellular automata (CA) technique \cite{ca, feature6}, whereby the movement of confined pedestrians is tracked over time. An important task related to evacuation simulations is how to describe the effects of subjective factors, such as fear, as well as the complicated forces among individuals. Helbing used ¡®social force¡¯ to simulate such interactions \cite{model2}, and Fukui avoided force concepts by applying the CA model \cite{ca}. Game theory was proposed to study the dynamics in crows evacuation \cite{game1, game2, game3, game4, game5, game6}. Heli$\ddot{o}$vaarawe presented a spatial game theoretic model for pedestrian behavior in situations of exit congestion \cite{game1}, and  Hao $etal$ applied game theory to deal with the conflicts that arise when two pedestrians try to occupy the same position\cite{game6}.

In this study, we modify the model \cite{game7} in which the effects of walls and defector herding are considered. A defector herding unit is defined as a group of four individuals who are all defectors in the same neighborhood. The potential well energy or mean force imposed on an individual is defined by considering three aspects. An individual will move to his selected location with a large probability if his personal energy is higher than the potential well energy. The velocity of a walker is defined as a function of the motion rule, and average velocity of the pedestrian system can reflect different dynamics, such as smooth flow, and congestion. We find that the panic level is crucial in determining the dynamics of the evacuation system and the competition between cooperators and defectors. It is also observed that three phases of pedestrian are exhibited sequentially as panic level upgrades.

The remainder of the paper is organized as follows. In Section II, we describe the model. Our numerical simulations and analysis are presented in Section III. Conclusions are drawn in Section IV.

\section{\label{sec:level1}model}

Pedestrian evacuation is studied on a model of square lattice on which all the individuals are distributed uniformly. Two pedestrian types are considered: cooperators, who adhere to the evacuation instructions; and defectors, who ignore the rules and act individually. Each pedestrian only interacts with his nearest neighbors, and is deleted from the system if he arrives at the exit. We apply game theory, and the payoff matrix representing the interactions among the crowd is shown in Table 1. To be specific, a cooperator receives a Reward ($R$) when he interacts with a collaborative neighbor, but suffers a Sucker ($S$) if he encounters a defector who gains a payoff of Temptation ($T$). Two defectors receive the Punishment ($PM$), respectively. In addition, a walker will gain a payoff $e$ ($0<e<R$) if there is an empty site in his neighborhood. Usually, $T=1+r$, $R=1$, $S=1-r$, $PM=0$, where $0<r<1$. $r$ is cost-to-benefit ratio indicating the defector-cooperator payoff divide, which describes the panic level during the escape. $e=R/2$ is set in this model.

\begin{table}[h]
\caption{Payoff matrix}
\begin{tabular}{@{\vrule height 10.5pt depth4pt  width0pt}lrcccc}
&\multicolumn5c{}\\
\noalign{\vskip-11pt}
  \\
\cline{2-6}
\vrule depth 6pt width 0pt {}&\multicolumn1c{$C$}&$D$&$E$&  \\
\hline
$C$&(R, R)&(S, T)&(e, $-$)& &\\
$D$&(T, S)&(PM, PM)&(e, $-$)& & \tablenote{C indicates an individual who cooperates, and D indicates an individual who chooses to defect. E represents an empty site. $T > R > S > PM$}
\\
\hline
\end{tabular}
\end{table}

$\overline{P}_{x}$ denotes the personal energy of individual $x$, and is the average of cumulative payoff that $x$ receives from all his neighbors. Each individual stays in a potential well which is formed by the neighboring people and walls (if walls exist in his neighborhood). $\overline{U}_{x}$ is the energy of potential well $x$ stays in, which consists of three parts. Firstly, $\overline{U}_{p}$ is the average cumulative payoff for all neighbors of $x$, which represents the mean forces on $x$:

\begin{equation}
\overline{U}_{p}=\frac{1}{G_{x}}\sum_{j\in G_{x}}\overline{P}_{xj}
\end{equation}
where $G_{x}$ is the scale of the interacting group centered on $x$, which is defined as the number of directly linked
neighbors of $x$ (excluding the empty sites). $\overline{P}_{xj}$ is the average of cumulative payoff that $j$ receives from all his neighbors, where
individual $j$ belongs to the neighborhood of $x$.

A \emph{defector herding unit} is defined as a group in which four defectors are in the same
neighborhood (Fig. 1). We consider that a defector herding unit contributes $\overline{P}_{D}=T/4$ to the energy
of the potential well for a cooperator, but contributes 0 to the energy of the potential well for a
defector. Therefore, the potential well energy $\overline{U}_{x}$ will include $\overline{U}_{du}$ if a defector herding unit exists
in a defector's neighborhood:

\begin{figure}
\centerline{\includegraphics[width=.4\textwidth]{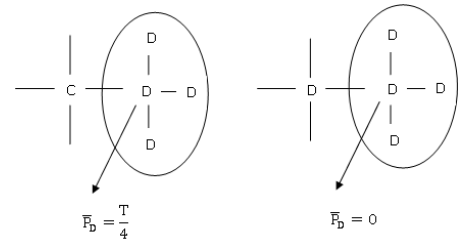}}
\caption{Schematic presentation of a defector herding unit in which all the individuals are defectors who are circled in the figure. $\overline{P}_{D}$ is the average payoff of the defector centered in the circle. }\label{afoto}
\end{figure}

\begin{equation}
\overline{U}_{du}=\frac{1}{G_{x}}(n_{du}\cdot \frac{T}{4})\cdot s,
\end{equation}
where $n_{du}$ is the number of defector herding units in the neighborhood of $x$. The parameter s takes the value $s = 0$ if $x$ chooses to cooperate, and $s = 1$ when $x$ decides to defect.

In addition, we consider the contribution of walls to $\overline{U}_{x}$, so $\overline{U}_{x}$ will include $\overline{U}_{w}$ if walls exist in the
neighborhood of $x$:
\begin{equation}
\overline{U}_{w}=\frac{1}{G_{x}}\cdot n_{w}\cdot U_{w},
\end{equation}
where $n_{w}$ is the number of walls in the neighborhood and $U_{w}$ $(0<U_{w}<R)$ is the contribution of a wall to $\overline{U}_{x}$.
Combining all the quantities discussed above, the potential well energy $\overline{U}_{x}$ is given by

\begin{equation}
\overline{U}_{x}=\overline{U}_{p}+\overline{U}_{du}+\overline{U}_{w},
\end{equation}
It should be noted that $x's$ personal energy is excluded in $U_{x}$ which actually reflects the pressures from $x's$ neighborhood.

At each time step, toward the exit's position, each individual $x$ randomly selects a cell $y$ in the neighborhood, and moves to $y$ with probability $W$ \cite{game7}:

\begin{equation}
W=\frac{1}{1+exp(-(\overline{P}_{x}-\overline{U}_{x})/\kappa)},
\end{equation}
where $\kappa$ is the noise amplitude. $\kappa=0$ indicates determined occupation, whereas $\kappa=\infty$ denotes stochastic occupation. It is apparent that the larger $\overline{P}_{x}$ is, the larger $W$ is. The equation (5) means that if $x$ owns more personal energy than the potential well energy, $x$ will move to his selected location with a larger probability. If $x$ takes the place of $y$, then the individual ever occupied $y$ will be pushed back to the original location of $x$.

Since equation [5] can reflect the motion velocity of an individual, $\overline{v}$ is defined as the system's average velocity and given by,

\begin{equation}
\overline{v}=\frac{1}{N_{M}}\sum_{j\in N_{M}}W_{j},
\end{equation}
where $N_{M}$ is the number of individuals who have the escape ability, and $W_{j}$ is the value of $W$ of individual $j$ representing its motion velocity. Different collective patterns of motion can be predicted according to $\overline{v}$: for $\overline{v}>0.5$, the system is realized to stay in the free flow; when $\overline{v}\rightarrow0$ the congestion forms. For $0<\overline{v}<0.5$, no distinct collective patterns emerge.

All the individuals will update their strategies synchronously by studying a more successful neighbor with a probability $W(s_{x}\rightarrow s_{y})$,
\begin{equation}
W(s_{x}\rightarrow s_{y})=\frac{1}{1+exp(-(\overline{P}_{y}-\overline{P}_{x})/\tau)},
\end{equation}
where $\tau$ is the noise level having the identical function of $\kappa$ in equation [5], and $s_{x}$ ($s_{y}$) is the strategy $x$ ($y$) adopts.

\section{\label{sec:level1}simulations and analysis}

The simulations are carried out on a square lattice representing a single room with a scale of $50\times 50$, and $2000$ pedestrians are considered. Only one exit exists, at a site range of $23-26$ on the lattice. Initially, all the pedestrians are distributed uniformly in the room with a cooperation density of $\rho_{IC}=0.5$. Escape is realized successfully when $95\%$ of the pedestrians have escaped from the room. We set $U_{w}=R/2$ and $\kappa=\tau=0.1$. In all of the simulations, each data point is the average for 100 realizations.

\begin{figure}
\centerline{\includegraphics[width=.4\textwidth]{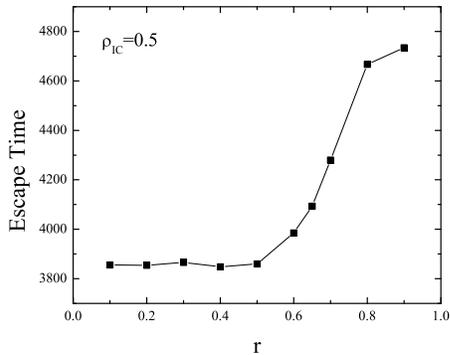}}
\caption{Escape time as a function of the panic level $r$ at the initial density of cooperators $\rho_{IC}=0.5$. }\label{afoto}
\end{figure}

\begin{figure}
\begin{center}
\scalebox{0.3}[0.3]{\includegraphics{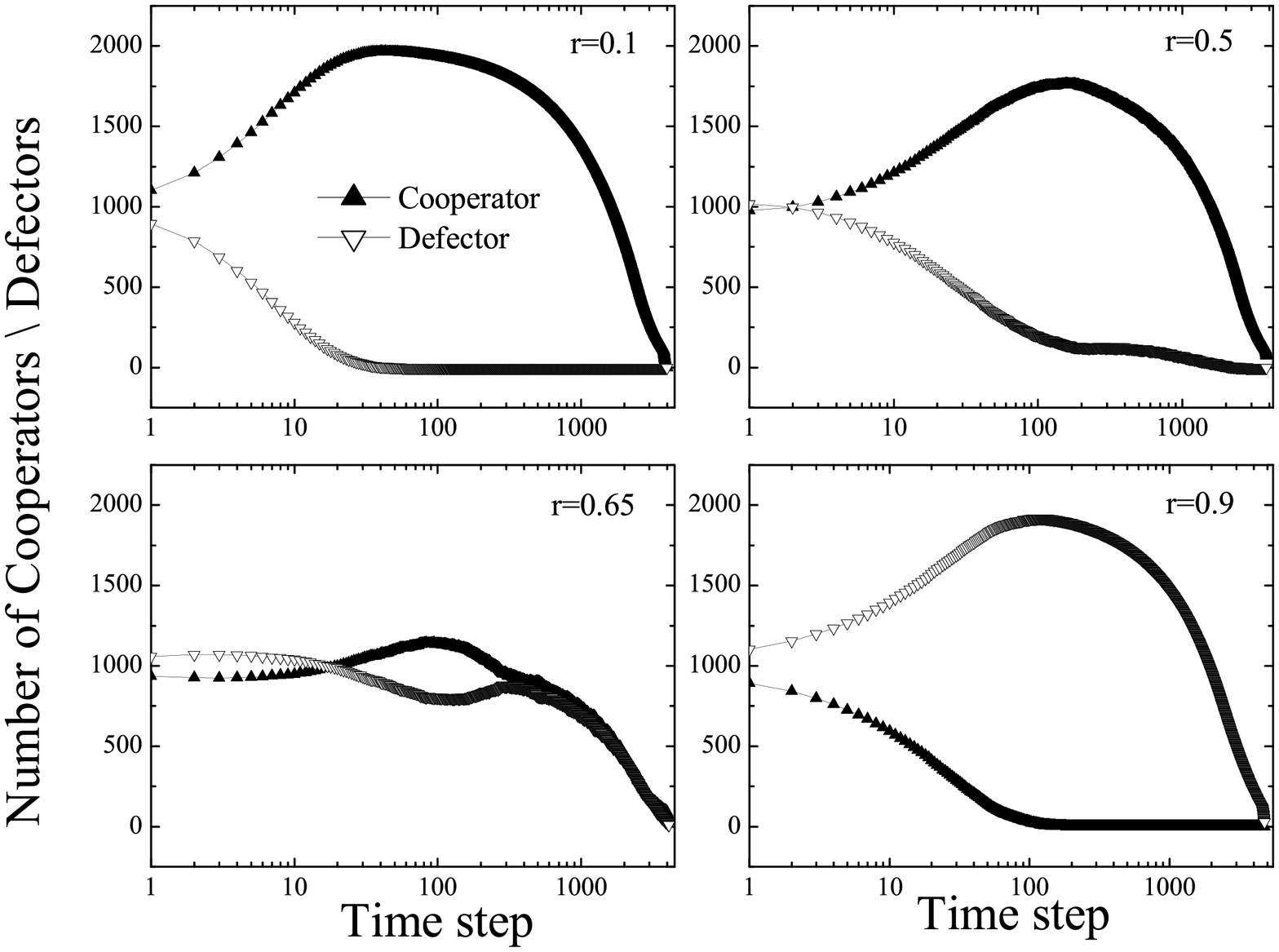}}
\quad
\scalebox{0.3}[0.3]{\includegraphics{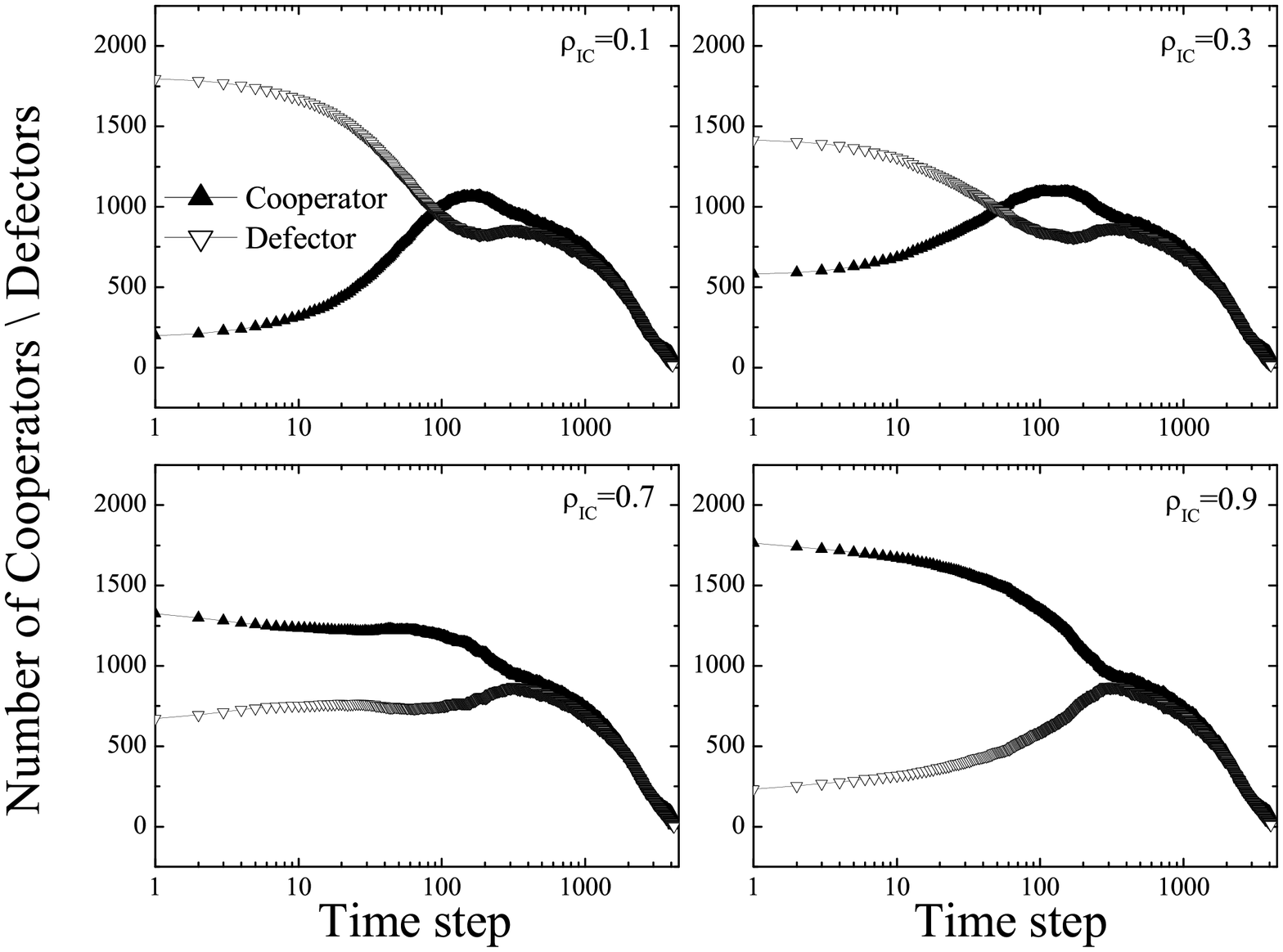}}
\caption{ Competition of cooperators and defectors for different panic levels of $r$ at $\rho_{IC}=0.5$ (UP), and Competition between cooperators and defectors for different values of $\rho_{IC}$ at $r = 0.65$ (DOWN).}
\end{center}
\end{figure}

\begin{figure}
\centerline{\includegraphics[width=.4\textwidth]{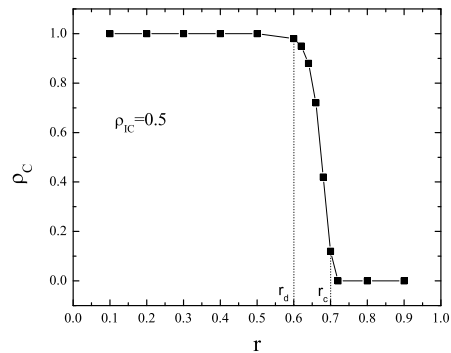}}
\caption{ Pedestrian phase transition with panic level $r$ when the initial cooperation density $\rho_{IC}=0.5$. $\rho_{C}$ on the vertical coordinate is cooperation density among the pedestrians in the room when 95\% of pedestrians have escaped out of the room. $r_{d}$ is the critical value at which defectors emerge, and $r_{c}$ is the critical value after which cooperators become extinct.}\label{afoto}
\end{figure}

Figure 2 shows the variation in escape time with panic level $r$ at $\rho_{IC}=0.5$. It is obviously seen that the escape time remains constant for small values of $r$ ($r < 0.5$), but increases with $r$ for $r\geq 0.5$. The results indicate that remaining calm is always the best strategy during escape, whereas too much fear leads to a longer escape time.

In order to understand the dynamics, interactions between cooperators and defectors are investigated. Different features are shown for different values of r (see it in Fig. 3 (UP)): (1) for $r = 0.1$ and $0.5$, cooperators dominate; (2) for $r = 0.65$, cooperators and defectors coexist and over time their frequencies reach almost identical levels; and (3) for $r = 0.9$, defectors dominate the system. In Fig. 3 (DOWN), it is noted that cooperators coexist with defectors, and evolve to eventually reach almost the same mutual frequency, regardless of $\rho_{IC}$. Combining this with the result in Fig. 3 (UP), we can conclude that the panic level $r$ remarkably affects the system dynamics, whereas $\rho_{IC}$ has no distinct impact.

The simulation results can be explained by analyzing the panic level. It is known from the payoff matrix that a high panic level induces a strong temptation to defect. According to the definitions of the potential well energy and motion rule in Equations $(1)-(5)$, a large increment in the percentage of defectors will lead to a relatively large increase in potential well energy or strong pressure, which accordingly reduces the escape velocity.

Moreover, pedestrian phase transition with panic level $r$ is discussed in Fig. 4. We observe that the system experiences three phases sequentially: a single phase of cooperator for $r < 0.6$, a mixed two-phase pedestrian of cooperator and defector for \textbf{$0.6(r_{d}) \leq r \leq 0.7(r_{c})$}, and a single phase of defector for $r > 0.7$. It has been proven that as $e$ increases from $0.1$ to $0.9$, $\overline{r_{d}}=0.6089$, the standard deviation is $\sigma(r_{d})=0.0105$; $\overline{r_{c}}=0.7044$, and $\sigma(r_{c})=0.0088$. As $U_{w}$ increases from $0.1$ to $0.9$, $\overline{r_{d}}=0.5956$, $\sigma(r_{d})=0.0133$; $\overline{r_{c}}=0.7111$, and $\sigma(r_{c})=0.0105$. It is obviously seen that the phase transition is basically robust to the changes of $e$ and $U_{w}$, which is consistent with the results shown in Fig. 4. Furthermore, noise effect $\kappa$ (in equation 5) was also studied: when $\kappa=0.001$ (approaching determined occupation), $r_{d}=0.58$, $r_{c}=0.7$; for $\kappa=100$ (approaching stochastic occupation), $r_{d}=0.54$, and $r_{c}=0.72$. It is concluded that phase transition is also basically robust to the noise, but the interval of mixed phase becomes broader when noise becomes very large.

To uncover the competition mechanism between the two types of pedestrians, mean payoffs of cooperators and defectors along the boundary are studied in Fig. 5. It is observed that in the phase of $r<0.6$, $\overline{P}_{c-bound}$ is significantly higher than $\overline{P}_{d-bound}$, whereas in the phase of $r >0.7$, $\overline{P}_{d-bound}$ is obviously higher. During the phase of $0.6 \leq r \leq 0.7$,  $\overline{P}_{d-bound}$ keeps a little higher, but approaches to $\overline{P}_{c-bound}$ as time evolves. Cooperator cluster is proven crucial in cooperation spread because of its firm boundary. According to the results in Fig. 5, it can be induced that cooperator cluster's boundary becomes weaker as panic level $r$ upgrades. Figure 6 shows the spatial evolution of pedestrians at $r=0.65$, in which the black symbols indicate the empty sites, the brown ones denote the defectors, and the white symbols represent cooperative pedestrians. It is seen that since defectors' boundary is slightly firmer than cooperators' (see Fig. 5 in the phase of $0.6 \leq r \leq 0.7$), many cooperator clusters are formed but don't spread all over the system.

\begin{figure}
\centerline{\includegraphics[width=.4\textwidth]{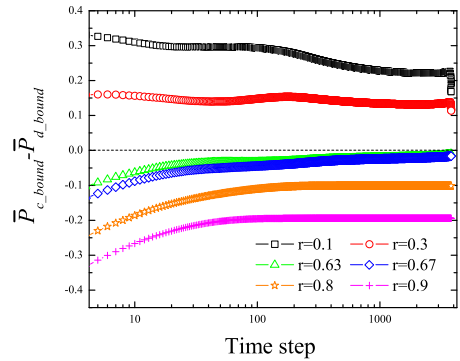}}
\caption{ (Color online) The differences between mean payoffs of cooperators and defectors along the boundary ($\overline{P}_{c-bound}-\overline{P}_{d-bound}$) at $\rho_{IC}=0.5$ for different panic levels of $r$. }\label{afoto}
\end{figure}

\begin{figure}
\centerline{\includegraphics[width=.4\textwidth]{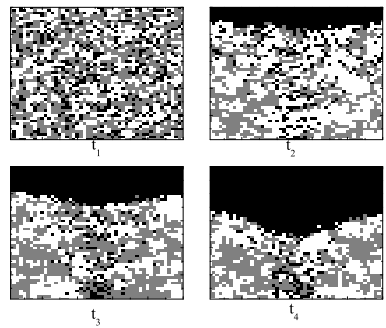}}
\caption{ (Color online) Spatial evolutions of cooperators and defectors at $\rho_{IC}=0.5$, and $r=0.65$. The black symbols represent empty sites, the brown ones indicate the defectors, and the white symbols denote cooperators. }\label{afoto}
\end{figure}

\begin{figure}
\centerline{\includegraphics[width=.4\textwidth]{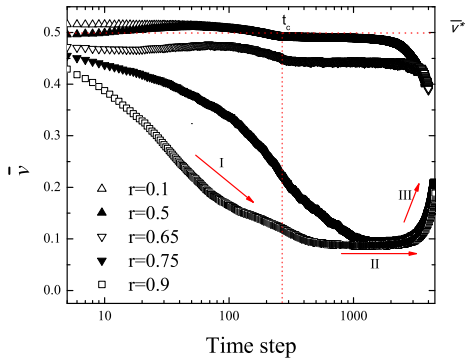}}
\caption{Evolution of average velocity $\overline{v}$ at $\rho_{IC}=0.5$ for different values of panic level $r$. $\overline{v}*=0.5$, $I$, $II$, and $III$ represent three stages a system experiences when $r >0.65$. }\label{afoto}
\end{figure}

The evolution of the average velocity $\overline{v}$ for different panic levels $r$ at $\rho_{IC}=0.5$ is shown in Fig. 7. It is seen that the average velocity decreases as the panic level increases, and declines sharply for $r> 0.65$. It is worthy noting that $\overline{v}$ changes slowly with $t$ for $r < 0.65$, but when $r > 0.65$, the velocity sharply decreases over time until the system enters a uniform motion state. This decrease in velocity is likely to be closely related to the emergence of congestion, and the uniform-motion state corresponds to free flow (moderate or relatively high velocity) or congestion (very low velocity). Therefore, from the phase aspect we can conclude that the system generally remains in a free-flow state for phase of $r < 0.6$ when $t<t_{c}$ ($\overline{v}>\overline{v}*$), and then moves at a relative low speed without distinct collective patterns. It exhibits three behavior stages for phase of $r > 7$: in stage I, local free flow and congestion coexist; in stage II, only congestion exists; and in stage III, congestion evacuation occurs. Congestion emerges before the system enters the uniform state owing to the small value for $\overline{v}$ in stage II. It can be predicted from this analysis that the probability of congestion is high for a high panic level. For the phase of coexistence ($0.6\leq r\leq7$), no distinct collective patterns happen.

Finally, Figure 8 shows the escape time as a function of $\rho_{IC}$ for $r = 0.6$. It is presented that pedestrians gain the greatest benefit for overall cooperation ($\rho_{IC}=1$), while the worst-case scenario occurs if all individuals defect ($\rho_{IC}=0$). The initial density of cooperators ($0<\rho_{IC}<1$) has no obvious influences on the escape time.

\begin{figure}
\centerline{\includegraphics[width=.4\textwidth]{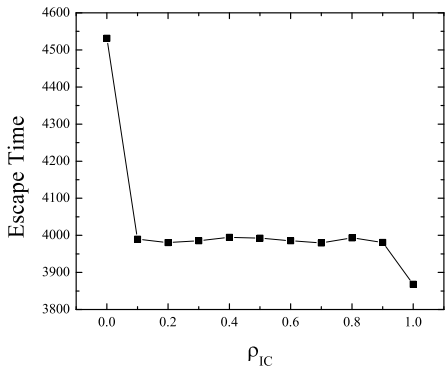}}
\caption{Variation of escape time with the initial density of cooperators $\rho_{IC}$ at $r=0.6$. }\label{afoto}
\end{figure}

\section{\label{sec:level1}Conclusions}

Extending the model proposed in previous work, we considered the effects of defector herding and walls to investigate pedestrian evacuation. Payoff represented the complicated interactions among individuals and was included in the energy representation. We found that the escape time increased with the panic level, and three phases of pedestrian were observed sequentially as panic level upgraded. Analysis of the results indicated that a high panic level induced a strong temptation to defect, which leaded to the emergence of widespread defection. Furthermore, a large percentage of defectors induced a comparatively large increase in potential well energy, which reduced the average speed of pedestrian flow according to the motion rule. Payoffs of cooperators and defectors along the boundary were investigated, and the results revealed the competition mechanisms between the two types of pedestrian. The average velocity was defined as the mean of cumulative motion probability in this model, which revealed valuable and interesting dynamics: the system was in the free flow when the panic level was low, but exhibited three stages - congestion formation, congestion, and congestion evacuation - for a high panic level. We proved that the panic level played an important part in determining the dynamics for which different behaviors were observed for different panic levels. Phase transition existed, and was basically robust to the changes of empty site contribution, wall's pressure and noise effect. We also found that global cooperation was the best strategy for the most efficient evacuation, and the situation was worst for all defections. It was proven that the initial density of cooperators had a negligible impact on the escape efficiency during evacuation.

\begin{acknowledgments}
This work was supported by Specialized Foundation for Theoretical Physics of China (Grant No. 11247239), National Natural Science Foundation of China (Grants No. 11305017, 11275186 and 91024026).
\end{acknowledgments}


\end{document}